\documentclass[preprint,aps]{revtex4}
\usepackage{graphicx}
\usepackage{hyperref}
\usepackage{epstopdf}
\usepackage{float}
\def\be{\begin{equation}}
\def\ee{\end{equation}}
\def\bea{\begin{eqnarray}}
\def\eea{\end{eqnarray}}
\def\ba{\begin{array}}
\def\ea{\end{array}}
\def\bc{\begin{center}}
\def\ec{\end{center}}

\begin{document}
\title{Generalized Parton Distributions of proton for non zero skewness in transverse and longitudinal position spaces}
\author{Narinder Kumar and Harleen Dahiya}
\affiliation{Department of Physics,\\ Dr. B.R. Ambedkar National
Institute of Technology,\\ Jalandhar, 144011, India}

\begin{abstract}
We investigate the Generalized Parton Distributions
(GPDs) of proton by expressing them in terms of overlaps of light front wave functions (LFWFs) using a simulated model which is able to qualitatively improve the convergence near the end points of $x$. We study the spin non-flip $H(x,  \zeta,  t)$  and spin flip $E(x,  \zeta, t)$ part of GPDs for the  particle conserving $n \rightarrow n$ overlap in the DGLAP region $(\zeta<x<1)$. The Fourier transform (FT) of the GPDs w.r.t. to the transverse momentum transfer as well the FT of the GPDs w.r.t. $\zeta$ have also been obtained giving the distribution of partons in the transverse position space and the distribution
in the longitudinal position space respectively. Diffraction pattern is obtained for both $ \mathcal{H}(x,\sigma,t)$ and $ \mathcal{E}(x,\sigma,t)$ in the longitudinal position space.

\end{abstract}
\maketitle

\section{Introduction}
The Compton scattering of coherent light on an object is one of the most elementary processes of physics. In a general way, by measuring the
angular and energy distributions of the scattered light information, the internal structure and the shape of the probed object can be accessed.
With the advent of intense multi-GeV lepton beam facilities, it has become possible to experimentally study the Compton scattering at the smallest dimensions of matter: the nucleon at quark and gluon level, where it is called Deep Virtual Compton scattering (DVCS).
The term virtual here has the meaning that the incoming photon is radiated from a lepton beam, which presents the additional advantage of varying its 3-momentum independently of its energy. DVCS process $\gamma^*(q)+p(P)\rightarrow \gamma(q')+ p(P')$, where the virtuality of the initial photon $q^2=-Q^2$ is much large compared to the squared momentum transfer $t=-(P-P')^2$, provides a valuable probe to the structure of the proton.

Within the framework of Quantum Chromodynamics (QCD), we intercept
such reaction at the partonic level through the concept of
Generalized Parton Distributions (GPDs) \cite{gpds1,gpds2,gpds3,gpds4}. They are experimentally accessed through the overlap of DVCS and Beith-Heitler process as well as exclusive vector meson production \cite{boffi}. Several experiments, such as, H1 collaboration \cite{2,3}, ZEUS collaboration \cite{4,5} and fixed target experiments at HERMES \cite{6} have finished taking data on DVCS. Experiments are also being done at JLAB, Hall A and B \cite{7} and COMPASS at CERN \cite{8}  to access GPDs.
GPDs not only allow us to access partonic configurations with a given longitudinal momentum fraction, similar to deep inelastic scattering (DIS), but also at specific (transverse) location inside the hadron. GPDs depend on three variables $x, \ \zeta, \ t$, where $x$ is the fraction of momentum transferred, $\zeta$ gives the longitudinal momentum transfer and $t$ is the square of the momentum transfer in the process. However, it has to be realized that only two of these variables $\zeta$ (fully defined by detecting the scattered lepton $\zeta=x_b$, where $x_b$ is the Bjorken variable used in DIS) and $t$ (fully defined by detecting either the recoil proton or meson) are accessible experimentally. GPDs are much richer in content  and carry much more information about the hadron structure than ordinary parton distributions. In the forward limit of zero momentum transfer, the GPDs reduce to ordinary parton distributions. The GPDs give interesting information about the spin and orbital angular momentum of the quarks and gluons in the nucleon.


It is well known that $\zeta$ represents the longitudinal momentum transfer in the DVCS process and in particular $\zeta= 0$ represents the momentum transfer only in the transverse direction. In Ref. \cite{burkardt1,burkardt2,burkardt3,pire}, GPDs have been studied for $\zeta=0$ and it has also been shown that Fourier transform of GPDs w.r.t $\Delta_\perp$ at $\zeta=0$ gives ipdpdfs (impact parameter dependent parton distribution functions). Since, in the DVCS process, the transformation of a virtual photon into real photon requires a finite transfer of longitudinal momentum and the  experiments always probe $\zeta \neq 0$, it becomes desirable to develop a deeper understanding of GPDs at $\zeta \neq 0$. The situation for $\zeta \ne 0$ also has an important difference from $\zeta=0$  because as the proton loses longitudinal momentum its transverse position is shifted by an amount which is proportional to $\zeta$ \cite{manohar}. Further, the “transverse position” of the proton is the vector sum of the transverse positions of its partons weighted by their longitudinal momentum fractions. Therefore, when the proton loses its momentum its transverse position get shifted from its original place.
Even though the polynomiality conditions have to be satisfied by the GPDs with $\zeta \neq 0$, it is very difficult to obtain a suitable parametrization of higher Fock states of the wave functions in certain models.  This is possible only if one considers the light front wave functions (LFWFs) of simple spin 1/2 objects like dressed quark or a dressed electron in theory.

GPDs can be expressed as overlaps of LCWFs of the target hadron in light front gauge \cite{brodsky1,kumar_dahiya1,kumar_dahiya2}.  We have generalized the
framework of QED by assigning a mass $M$ to external electrons in the Compton scattering process, but a different mass $m$ to the internal electron
line and a mass $\lambda$ to the internal
photon line \cite{brodsky2}. The idea behind this is to model the
structure of a composite fermion state with a mass $M$ by a fermion and a vector constituent with respective masses $m$ and $\lambda$.
These are off-forward overlaps in general and one requires not only particle conserving $n\rightarrow n$ overlap similar to forward pdfs
but also $n+1\rightarrow n-1$ overlap. Therefore, a spin 1/2 system can be represented as a composite of a spin 1/2 fermion and spin 1 vector boson with arbitrary masses \cite{brodsky3,brodsky4}.
This one loop model is self consistent since it has correct correlation of different Fock components of the state as given by light front eigen value equation. This model has been used to calculate the spin and orbital angular momentum of a composite relativistic system as well as the GPDs in impact parameter space and it also gives the Schwinger anomalous magnetic moment and the corresponding Dirac's and Pauli form factor including the vanishing of the anomalous gravitomagnetic moment B(0). It provides a template for the wave function of an effective quark-diquark model of the valence Fock state of the proton LFWF. We would like to emphasize here that this is extremely difficult to achieve in phenomenological models. In particular, two or three body Fock components can be obtained analytically from QED. Further, by taking the Fourier conjugate $b_\perp$ (impact parameter) of the transverse momentum transfer $\Delta_\perp$, the GPDs can be expressed in impact parameter space.

In Ref. \cite{brodsky3,brodsky4}, the model with simulated bound state has been discussed. It is clear from these calculations that without involving the simulated bound state of the wave functions, one has to introduce the cut off on the momenta in the integrals for the expressions of GPDs. But differentiating the wave function w.r.t bound state mass $M^2$ improves the behavior of wave functions near the end points of $x$ as well as at high $k^2_\perp$. Differentiating  the wave function w.r.t. $M^2$ produces a meson like behavior but it does not represent a model for meson. Even though some work has been done in the simulated model of LFWFs \cite{15} where GPDs were discussed in impact parameter space i.e. by considering $\zeta = 0$ but no work has been done so far for the case of $\zeta \ne 0$.
There are three distinct regions in $x$. In the domain where
$\zeta<x<1$, there are diagonal $2\rightarrow 2$ overlap contributions for both helicity non-flip $H_{2\rightarrow 2}(x,  \zeta,  t)$
and helicity flip $E_{2\rightarrow 2}(x,  \zeta,  t)$. The GPDs $H_{2\rightarrow 2}(x,\zeta,t)$ and $E_{2\rightarrow
2}(x,\zeta,t)$ vanishes in the domain $\zeta -1 <x<0$. In the domain $0<x<\zeta$, we have contribution from non diagonal $n+1\rightarrow n-1$ overlap. Since the differentiation of the single particle LFWF gives a vanishing result, this contribution vanishes in this model. It has also been shown that the DVCS amplitude expressed in terms of the variable $\sigma$ show diffraction pattern analogous to diffractive scattering of a wave in optics where the distribution in $\sigma$ measures the physical size of the scattering center in a 1-D system \cite{brodsky3,brodsky4,14}. The finite size of the $\zeta$ integration of the Fourier transform acts as a slit width and produces the diffraction pattern. Therefore, it becomes interesting to study the GPDs and the Fourier transform (FT) of the GPDs in the transverse as well as longitudinal position space.

In the present work, we have calculated the $H$ and $E$ for non zero skewness. To understand the significance of the momentum fraction carried by the quarks in the process, we have considered the cases for fixed $x$ and different values of $\zeta$ as well as for fixed $\zeta$ and different values of $x$. Both GPDs $H$ and $E$ are studied in longitudinal and transverse position space by taking Fourier transform w.r.t. $\zeta$ and $\Delta_\perp$ giving the distribution of partons in the longitudinal position space and the distribution
in the transverse position space respectively.

\section{Kinematics Of DVCS and Generalized Parton Distributions}

For the sake of completeness, we present the essential kinematics of deeply virtual Compton scattering \cite{brodsky1}
\be
\gamma^*(q) + p(P) \rightarrow \gamma(q') +p(P')\,.
\ee
The frame is specified by choosing a convenient parametrization of the light-cone coordinates for the initial and final proton:
\bea
P &=& \left(P^+,\ \vec{0}_\perp, \ \frac{M^2}{P^+} \right) \,,
\\
P^{'} &=& \left( (1-\zeta)P^{+},\ -\vec{\Delta}_\perp,\
\frac{M^{2}+ {\vec \Delta}^2_\perp }{(1-\zeta)P^{+}} \right) \,,
\eea
where $M$ is the proton mass. The four momentum transfer from the target is
\begin{eqnarray}
\Delta =P-P^{'}= \left(\zeta P^{+},\ \vec{\Delta}_\perp,\ \frac{t+ {\vec \Delta}^2_\perp }{\zeta P^{+}} \right)\,,
\label{del}
\end{eqnarray}
where  $ t= \Delta^2$. In addition, overall energy-momentum conservation requires $\Delta^{-}=P^{-} - P'^{-}$, which connects ${\vec \Delta}^2_\perp$, $\zeta$ and $t$ as follows
\be
t=2 P \cdot \Delta=-\frac{\zeta^{2} M^2+\vec{\Delta}^2_\perp}{1-\zeta} \,.
\ee
The generalized form factors $H$, $E$ are defined through the matrix elements of the bilinear vector currents on the light cone \cite{brodsky1}:
\begin{eqnarray}
\lefteqn{\int{\frac{dy^-}{8\pi}} e^{i x P^+ y^-/2} \langle P'|\bar{\psi}(0)\gamma^{+} \psi(y)|P\rangle|_{y^+=0, \ y^\perp=0} } \nonumber\\
&&=\frac{1}{2 \bar{P^+}} \bar{U}(P')[H(x,\zeta,t)\gamma^+ + E(x,\zeta,t) \frac{}{2M}\sigma^{+\alpha}(-\Delta_\alpha)]U(P) .
\end{eqnarray}
The off-forward matrix elements can be expressed as overlaps of the light front wave functions \cite{13}. For non-zero skewness $\zeta$ there are diagonal parton number conserving contributions in the kinematical region $\zeta<x<1$ and $\zeta-1<x<0$ and there are parton number changing contributions in the region $0<x<\zeta$.
If we consider a spin 1/2 target state consisting of a spin 1 particle and a spin 1/2 particle, the contribution to the spin flip conserving and non conserving part of GPDs in the domain $\zeta<x<1$ can be expressed as
\[\frac{\sqrt{1-\zeta}}{1-\frac{\zeta}{2}}H_{2\rightarrow2}
(x,\zeta,t)-\frac{\zeta^2}{4(1-\frac{\zeta}{2}) \sqrt{1-\zeta}}E_{2\rightarrow2}(x,\zeta,t) = \]
\begin{equation} \int{\frac{d^2\vec{k}_\perp}{16\pi^3}} \bigg[ \psi_{+\frac{1}{2}+1}^{\uparrow *}(x',\vec{k}'_\perp) \psi_{+\frac{1}{2}+1}^{\uparrow}(x,\vec{k}_\perp)+\psi_{+\frac{1}{2}-1}^{\uparrow *}(x',\vec{k}'_\perp) \psi_{+\frac{1}{2}-1}^{\uparrow}(x,\vec{k}_\perp)+
\psi_{-\frac{1}{2}+1}^{\uparrow *}(x',\vec{k}'_\perp) \psi_{-\frac{1}{2}+1}^{\uparrow}(x,\vec{k}_\perp) \bigg] \,,
\label{h22}
\end{equation}
\begin{eqnarray}
\lefteqn{ \frac{1}{\sqrt{1-\zeta}}\frac{\Delta^1-i \Delta^2}{2M}E_{2\rightarrow2}(x,\zeta,t)=\nonumber}\\
&&\int{\frac{d^2\vec{k}_\perp}{16\pi^3}} \Big[\psi_{+\frac{1}{2}+1}^{\uparrow *}(x',\vec{k}'_\perp) \psi_{+\frac{1}{2}+1}^{\downarrow}(x,\vec{k}_\perp)+\psi_{+\frac{1}{2}-1}^{\uparrow *}(x',\vec{k}'_\perp) \psi_{+\frac{1}{2}-1}^{\downarrow}(x,\vec{k}_\perp) \Big]\,,
\label{e22}\end{eqnarray}
where
\be
x'=\frac{x-\zeta}{1-\zeta}, ~~
\vec{k}'_\perp=\vec{k}_\perp-\frac{1-x}{1-\zeta}\vec{\Delta}_\perp\,.
\ee
\label{sim}
We calculate the GPDs in simulated model of hadron LFWFs. To begin with, we present the two-particle wave function for spin up and spin down electron, which can be expressed as \cite{13}:
\begin{eqnarray}
&&\psi_{+\frac{1}{2}+1}^{\uparrow}(x,\vec{k}_\perp)=-\sqrt{2}\frac{-k^1+i k^2}{x(1-x)}\varphi,\nonumber\\ && \psi_{+\frac{1}{2}-1}^{\uparrow}(x,\vec{k}_\perp)=-\sqrt{2}\frac{k^1+i k^2}{(1-x)}\varphi,\nonumber\\ &&
\psi_{-\frac{1}{2}+1}^{\uparrow}(x,\vec{k}_\perp)=-\sqrt{2}(M-\frac{m}{x})\varphi,\nonumber\\ &&\psi_{-\frac{1}{2}-1}^{\uparrow}(x,\vec{k}_\perp)=0 \,,
\end{eqnarray}
and
\begin{eqnarray}
&&\psi_{+\frac{1}{2}+1}^{\downarrow}(x,\vec{k}_\perp)=0,\nonumber\\&&
\psi_{+\frac{1}{2}-1}^{\downarrow}(x,\vec{k}_\perp)=-\sqrt{2}(M-\frac{m}{x})\varphi,\nonumber\\ &&\psi_{-\frac{1}{2}+1}^{\downarrow}(x,\vec{k}_\perp)=-\sqrt{2}\frac{-k^1+i k^2}{(1-x)}\varphi,\nonumber\\
&&\psi_{-\frac{1}{2}-1}^{\downarrow}(x,\vec{k}_\perp)=-\sqrt{2}\frac{k^1+i k^2}{x(1-x)}\varphi \,,
\end{eqnarray}
where
\begin{eqnarray}
\varphi(x, \vec{k}_{\perp}) =\frac{e}{\sqrt {1-x}} \frac{1}{M^2-\frac{\vec{k}_{\perp}^{2}+m^2}{x}-\frac{\vec{k}_{\perp}^{2}+\lambda^2}{1-x}}\,.
\end{eqnarray}
As discussed earlier, we have used the generalized form of QED by assigning a mass $M$
to the external electrons and a different mass $m$ to the internal
electron lines and a mass $\lambda$ to the internal photon lines.
\begin{figure}
\minipage{0.42\textwidth}
    \includegraphics[width=5.8cm ,angle=270]{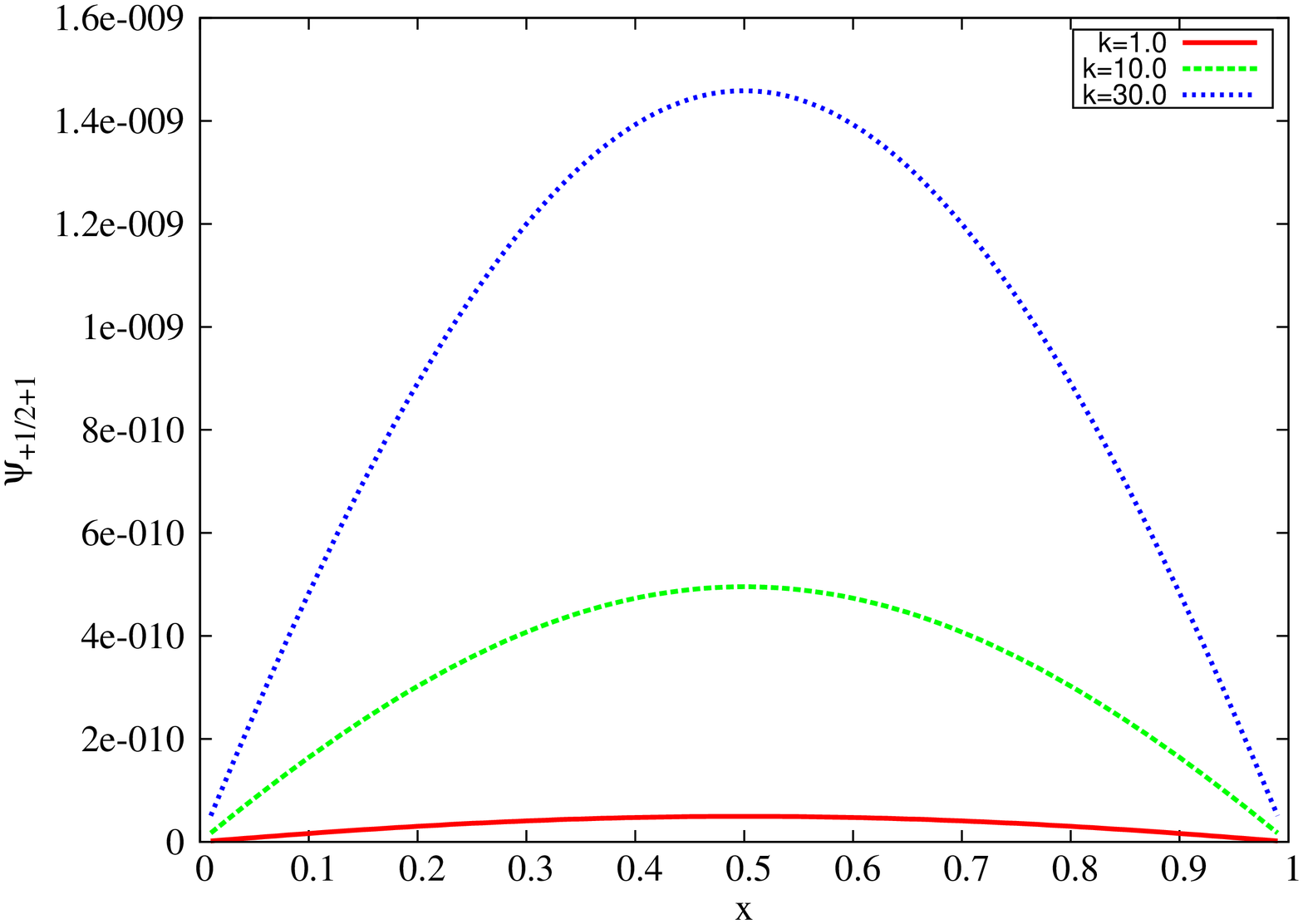}
  \endminipage\hfill
  \minipage{0.42\textwidth}
  \includegraphics[width=5.8cm ,angle=270]{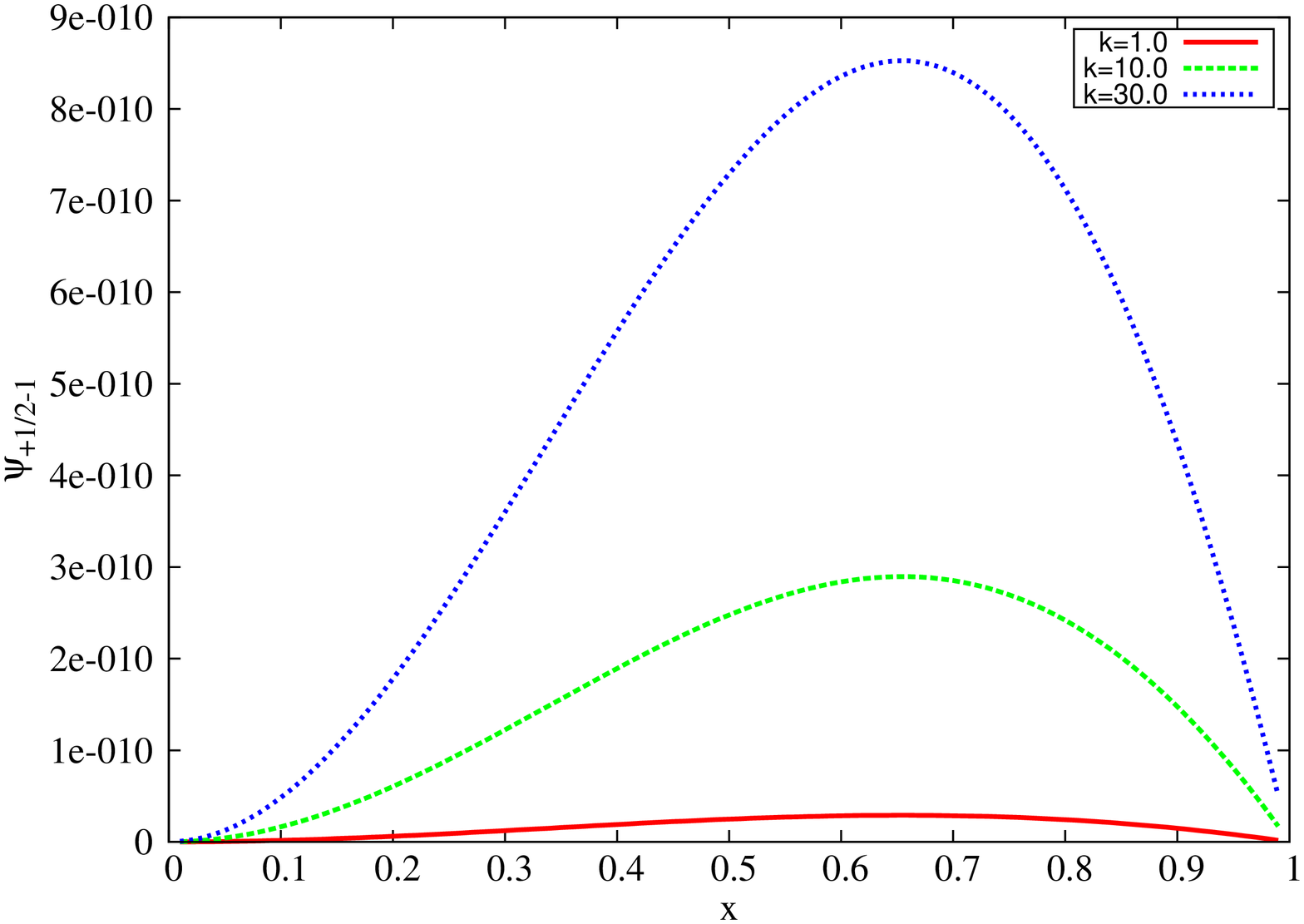}
\endminipage\hfill
  \minipage{0.42\textwidth}
  \includegraphics[width=5.8cm,angle=270]{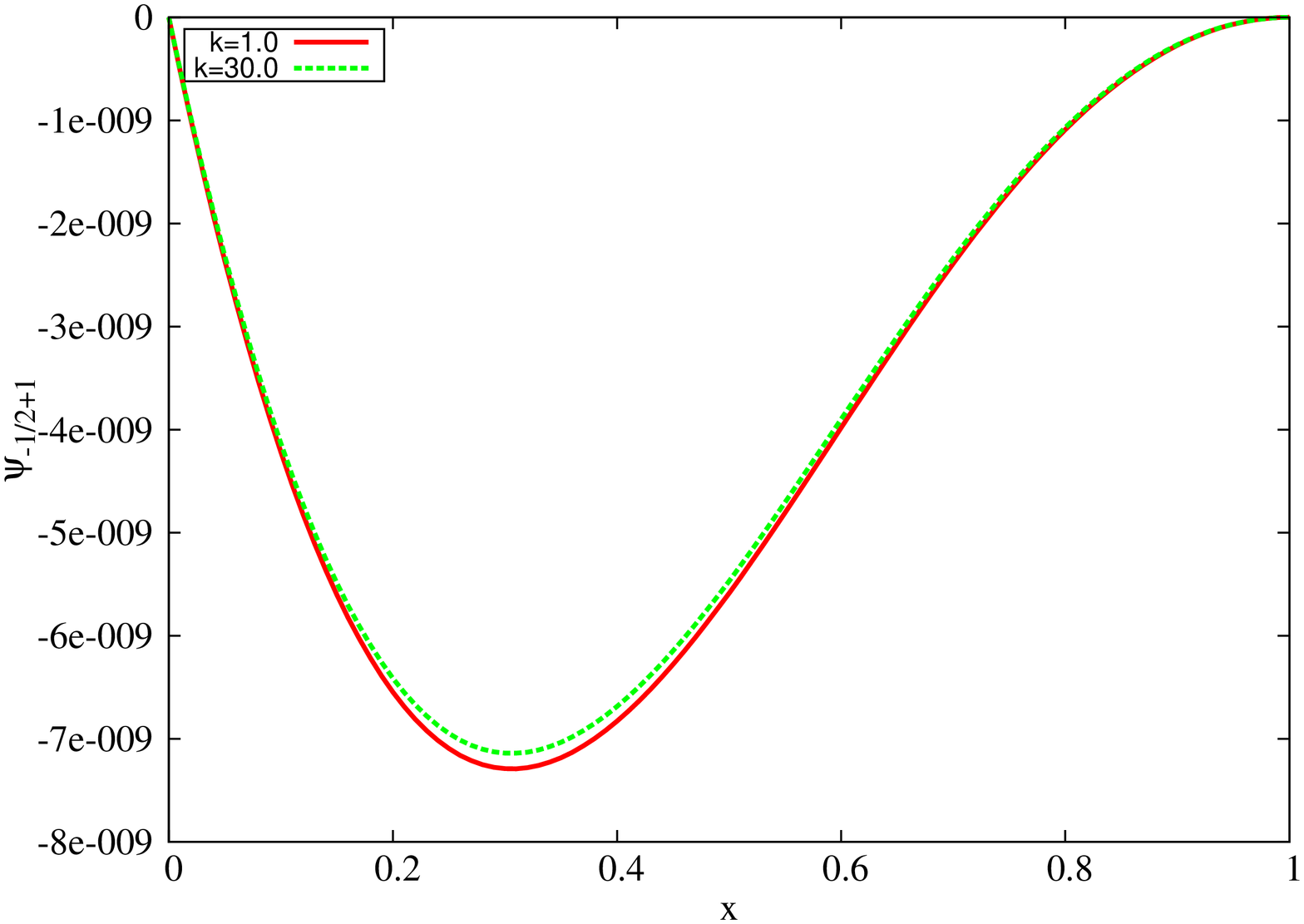}
  \endminipage\hfill
  \caption{Two-particle LFWFs vs $x$ for $M=150$ MeV, $m=300$ MeV,
$\lambda=300$ MeV and fixed values of $| k^{\perp}|=k$ in units of MeV.}
\label{wf}
\end{figure}

Differentiating the QED LFWFs with respect to $M^2$ improves the
convergence of the wave functions at the end points of $x$ as well as the $k_{\perp}^2$ behavior.
We differentiate $\varphi(x,\vec{k}_{\perp})$ with respect to $M^2$ and have
\begin{equation}
\varphi'(x,\vec{k}_{\perp})= \frac{e}{\sqrt{1-x}} \frac{1}{{
\left(M^2-\frac{\vec{k}_\perp^2 +m^2}{x}-\frac{\vec{k}_\perp^2
+\lambda^2}{1-x} \right)}^2} \,.
\end{equation}
Using this, in Fig.\ref{wf} we have plotted the two particle
LFWFs as a function of $x$ for different values of $k$. It is clear from the plots that the wave functions converge at the end points of $x$ and it's behaviour is improved.

The helicity non-flip GPD $H$ can be calculated in this model, using Eqs. (7) and (8), and can be expressed as
\bea
&&H(x,\zeta,t)=\frac{1-\zeta /2}{\sqrt{1-\zeta}} \left(\frac{x(x-\zeta)(1-x)}{(1-\zeta)^\frac{3}{2}}+\frac{x^2 (1-x)(x-\zeta)^2}{(1-\zeta)^\frac{5}{2}} \right)
\Big[I_1+I_2+ B I_3+\nonumber\\
&&\frac{2 (M(x-\zeta)-m(1-\zeta))(M x-m)x(1-x)^3(x-\zeta)}{(1-\zeta)^\frac{7}{2}} I_3 \Big] +
\frac{\zeta^2}{4 (1-\zeta)} E(x,\zeta,t)\,.
\eea

The helicity flip term $E$ can similarly be calculated from Eq. (\ref{e22}) and is given as
\bea
&& E(x,\zeta,t) = 4 M \sqrt{1-\zeta} \bigg(\bigg(\frac{(1-\zeta)}{1-x}\frac{(M x-m)}{x}-\frac{1}{1-x}\bigg(\frac{M (x-\zeta)-m(1-\zeta)}{x-\zeta}\bigg)\bigg) I_4 \nonumber\\
&&-(M-\frac{m}{x}) I_3 \bigg) \frac{x^2 (1-x)^4 (x-\zeta)^4}{(1-\zeta)^2}\,,
\eea
where
\bea
I_1 &=& \int{\frac{{\mathrm d}^2 k_\perp}{L_1 L_{2}^2}}=\pi \int_{0}^{1}{\frac{(1-\alpha) {\mathrm d} \alpha}{D^2}},
\nonumber\\
I_2 &=& \int{\frac{{\mathrm d}^2 k_\perp}{L_{1}^2 L_2}}=\pi \int_{0}^{1}{\frac{\alpha {\mathrm d} \alpha}{D^2}},
\nonumber\\
I_3 &=& \int{\frac{ {\mathrm d} ^2 k_\perp}{L_{1}^2 L_{2}^2}}=\pi \int_{0}^{1}{\frac{\alpha (1-\alpha) {\mathrm d} \alpha}{D^3}},
\nonumber\\
I_4 &=& \int{\frac{{\mathrm d}^2 k_\perp \vec{k}_\perp}{L_1^2 L_2^2}}=2 \pi \frac{1-x}{1-\zeta} \int_{0}^{1} \frac{y (1-y)^2 \Delta_\perp {\mathrm d} y}{Q^3},
\eea
\bea
L_1 &=& k_{\perp}^2-M^2 x (1-x)+m^2(1-x)+\lambda^2 x,\
\nonumber\\
L_2 &=& k_{\perp}^2-\frac{M^2(x-\zeta)(1-x)}{(1-\zeta)^2}-
2 \frac{(1-x)}{1-\zeta} k_{\perp} \cdot \Delta_{\perp}+
\frac{(1-x)^2 \Delta_{\perp}^2}{(1-\zeta)^2} +\frac{m^2(1-x)}{(1-\zeta)}+
\nonumber\\
&& \frac{\lambda^2 (x-\zeta)}{1-\zeta},
\nonumber\\
D &=& \alpha (1-\alpha)\frac{(1-x)^2 \Delta_{\perp}^2}{(1-\zeta)^2}+ m^2 (1-\alpha)
(1-x)+ \lambda^2 (1-\alpha) x -M^2 (1-\alpha) x (1-x)-\nonumber\\
&&\frac{M^2 \alpha (x-\zeta)(1-x)}{(1-\zeta)^2}+\frac{ m^2 \alpha (1-x)}{(1-\zeta)}+\frac{\lambda^2 \alpha (x-\zeta)}{1-\zeta} \,,
\nonumber\\
Q &=& -\frac{(1-y)^2 (1-x)^2 \Delta_{\perp}^2}{(1-\zeta)^2} + (1-y) (\frac{(1-x)^2 \Delta_{\perp}^2}{(1-\zeta)^2}- \frac{M^2 (x-\zeta) (1-x)} {(1-\zeta)^2} + \frac{m^2 (1-x)}{1-\zeta}+\nonumber\\
&& \frac{\lambda^2 (x-\zeta)}{1-\zeta} ) +y (m^2 (1-x) - M^2 x (1-x)+\lambda^2 x),
\nonumber\\
B &=& M^2 x (1-x)-\frac{(1-x)^2 \Delta_{\perp}^2}{(1-\zeta)^2}-
m^2(1-x)-\lambda^2 x+\frac{M^2(x-\zeta)(1-x)}{(1-\zeta)^2}-
\frac{m^2(1-x)}{1-\zeta}- \nonumber\\
&& \frac{\lambda^2 (x-\zeta)}{1-\zeta}.
\eea
\begin{figure}
\minipage{0.42\textwidth}
    \includegraphics[width=5.8cm,angle=270]{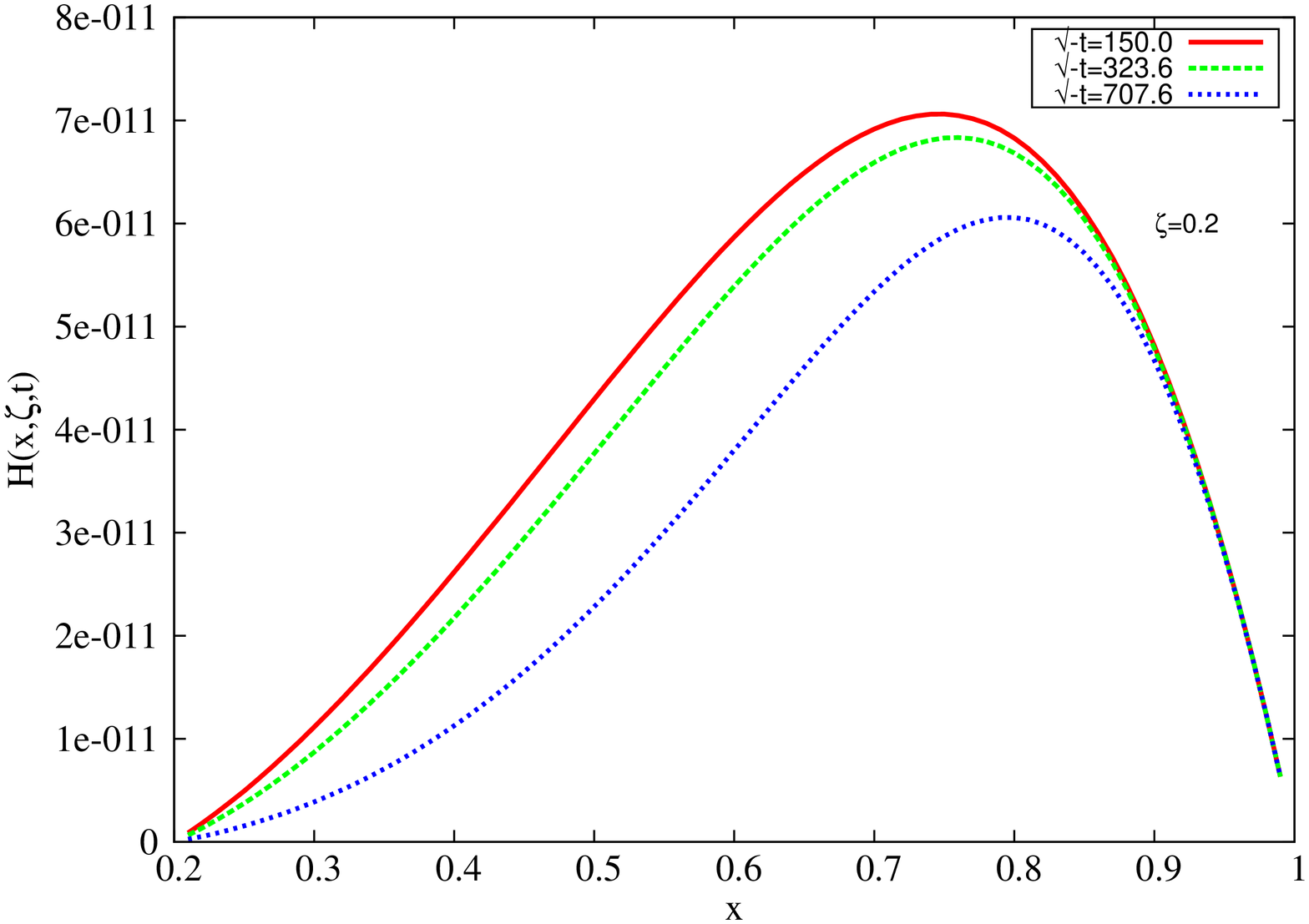}
  \endminipage\hfill
  \minipage{0.42\textwidth}
  \includegraphics[width=5.8cm,angle=270]{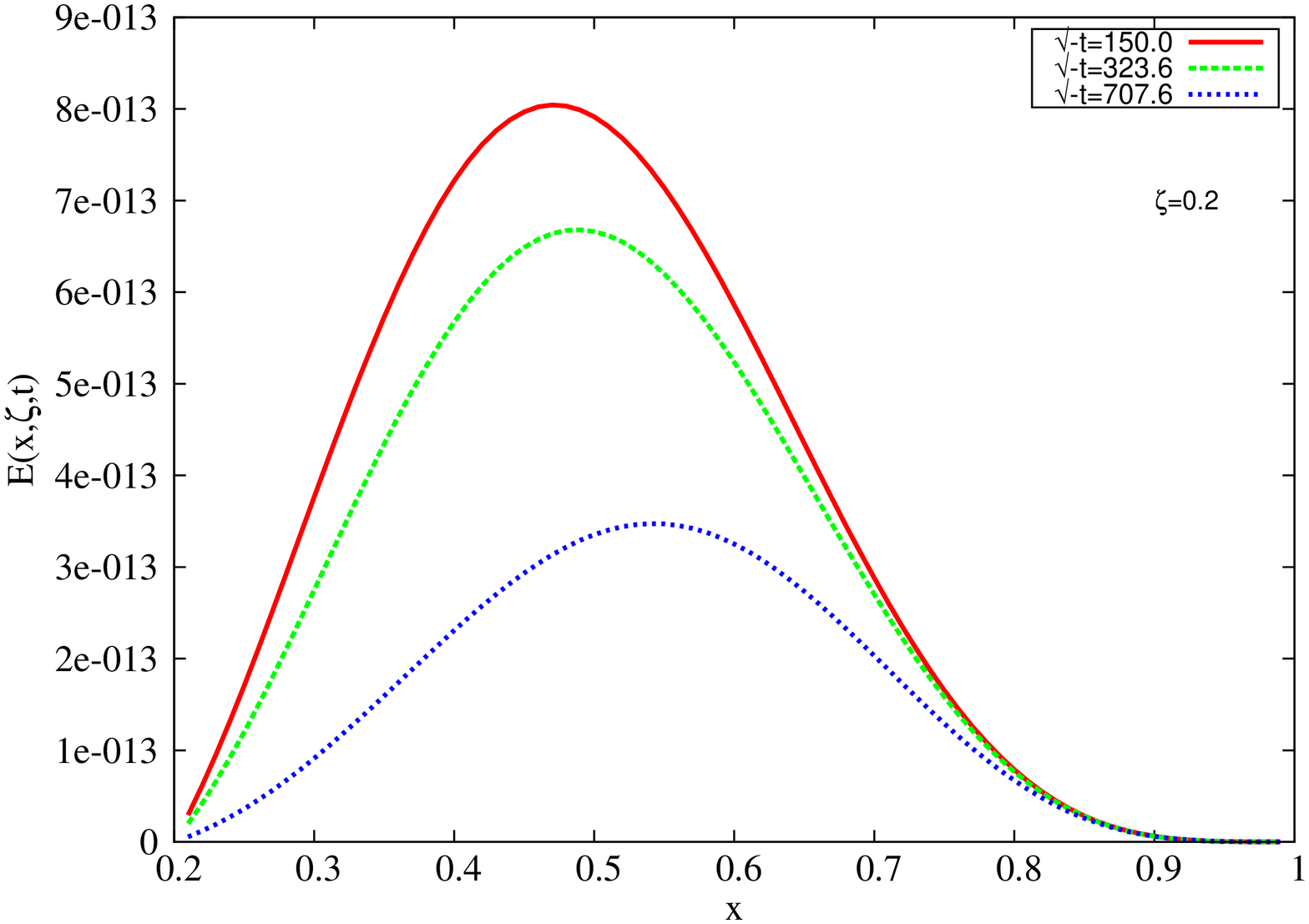}
  \endminipage\hfill
  \caption{Plot of $H$ vs $x$ for a fixed value of $\zeta$
with different values of $-t$ and plot of $E$ vs
$x$ for a fixed value of $\zeta$ with different values of $-t$. Parameter $t$ is in ${MeV}^2$}.
\label{newx}
\end{figure}
\begin{figure}[!]
\minipage{0.42\textwidth}
   \includegraphics[width=5.8cm,angle=270]{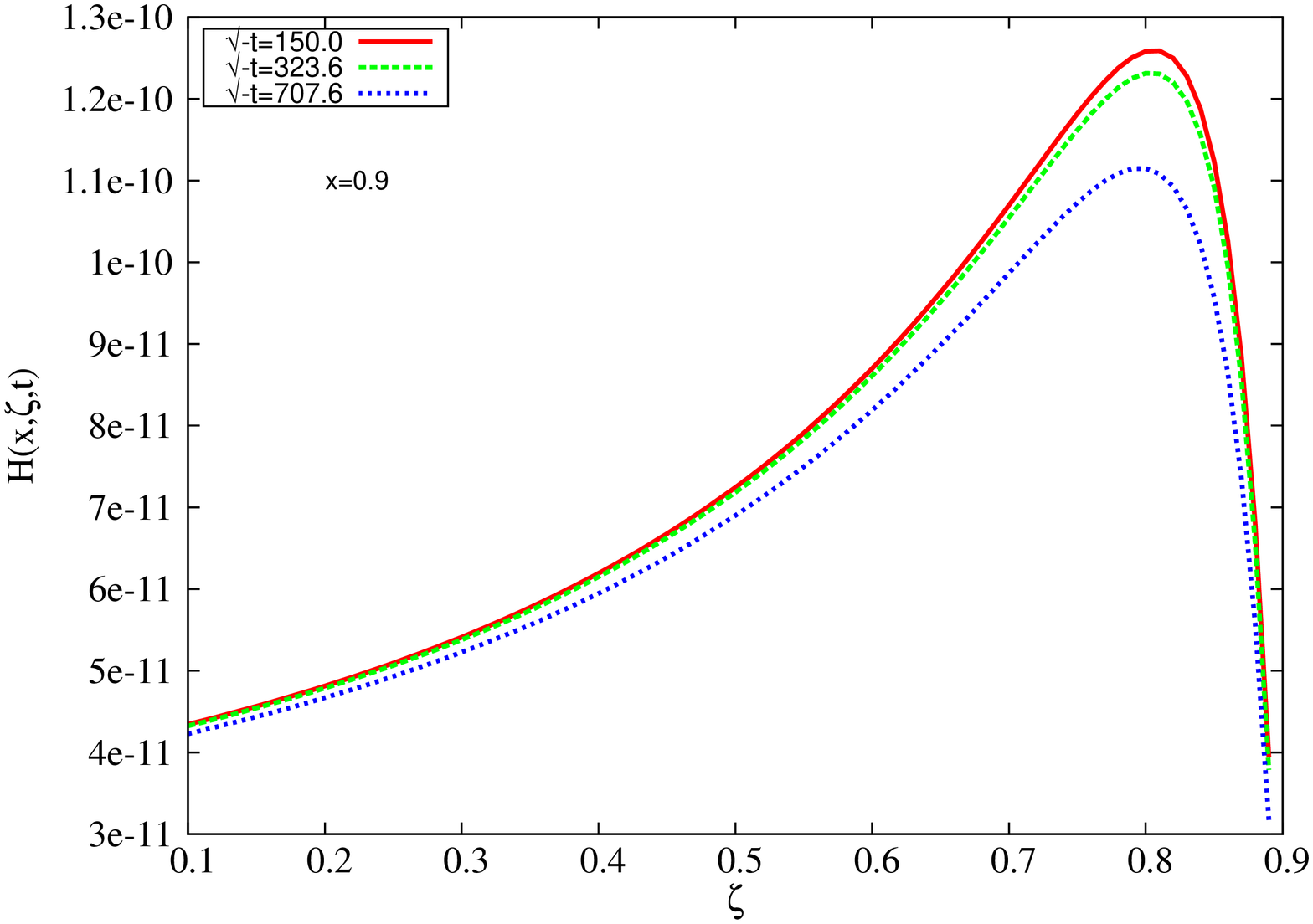}
  \endminipage\hfill
  \minipage{0.42\textwidth}
  \includegraphics[width=5.8cm,angle=270]{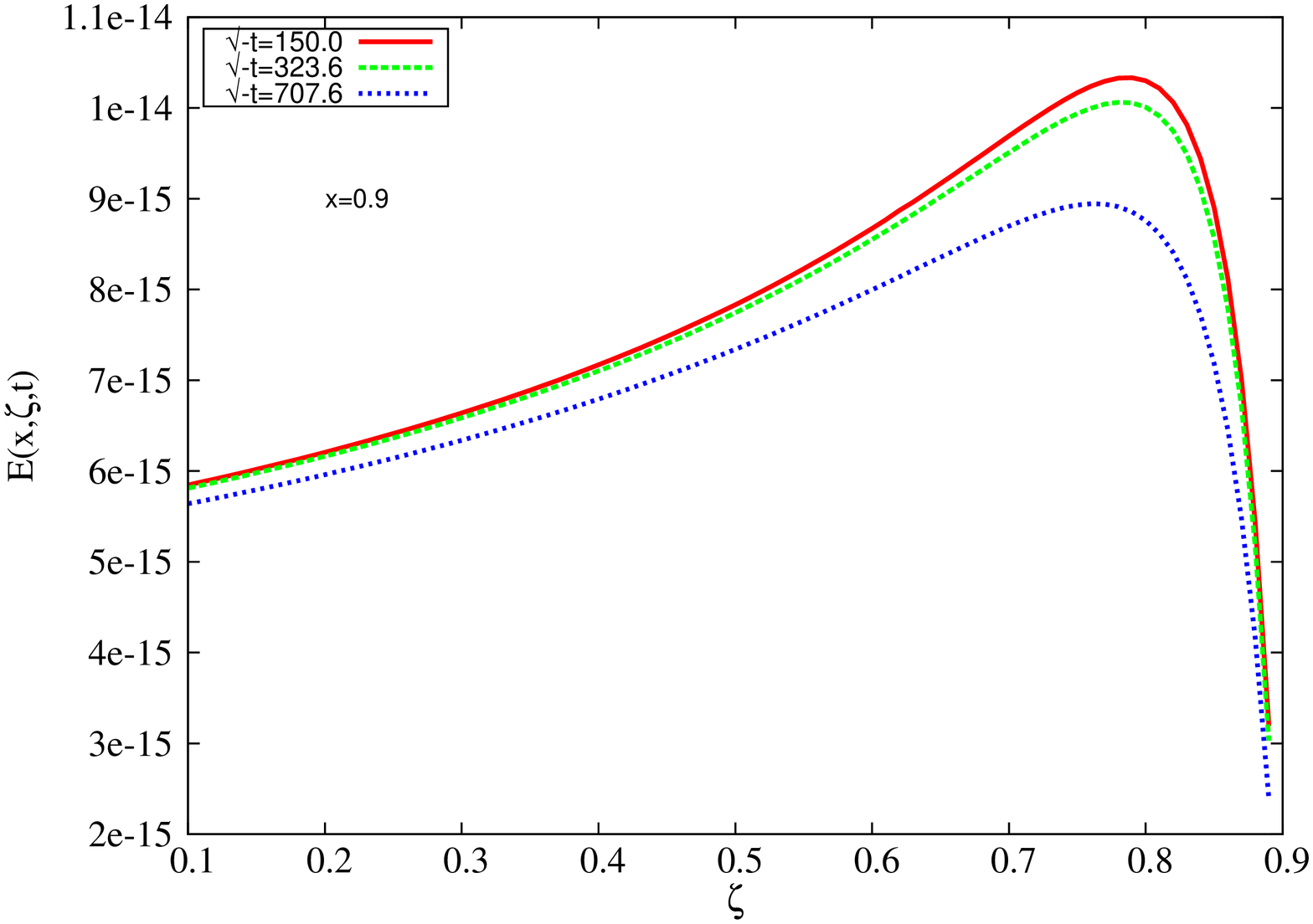}
  \endminipage\hfill
   \caption{Plot of $H$ vs $\zeta$ for a fixed value
of $x$ with different values of $-t$ and plot of  $E$ vs $\zeta$ for a fixed value of $x$ with different values of $-t$. Parameter $t$ is in  $MeV^2$}.
\label{newz}
\end{figure}
To understand the behaviour of the GPDs with the change in momentum transfer, in Fig. \ref{newx}, we have plotted the helicity non-flip GPD $H(x,\zeta,t)$ and helicity flip GPD $E(x,\zeta,t)$ as a function of $x$ with fixed value of $\zeta$ whereas in Fig. \ref{newz} we have plotted the helicity non-flip GPD $H(x,\zeta,t)$ and helicity flip GPD $E(x,\zeta,t)$ as a function of $\zeta$ with fixed value of $x$. For the numerical calculations we have used
$m =300$ MeV, $M =150$ MeV, and $\lambda =300$ MeV \cite{brodsky3,brodsky4,15}. A cursory look at the plots reveal that the behaviour of the GPDs is the same and is independent of the value of momentum transfer $|t|$. In Fig. \ref{newx},  $H(x,\zeta,t)$ and $E(x,\zeta,t)$ increase with $x$, reach maximum and then decrease. It is important to mention here that since $x$ is the momentum fraction of the active quark, at $x=1$, the active quark carries all the momentum and the contribution from other partons is expected to be zero at this limit. It is also observed that the peak of $H(x,\zeta,t)$ shifts towards higher value of $x$ as $|t|$ increases suggesting that the active quark is more likely to have a large momentum fraction. We are considering here only the leading Fock space component of the target hadron state. The higher Fock space components are more likely to contribute in the small $x$ region.  The helicity flip GPD $E(x,\zeta,t)$, related with the quark orbital angular momentum,  also decreases as $|t|$ increases. In the case, the peak occurs at a lower value of $x$ as compared to the peak in the case of $H(x,\zeta,t)$. However, the shifting of peak towards higher value of $x$ as $|t|$ increases is similar to that in the case of $H(x,\zeta,t)$.

On the other hand, in Fig. \ref{newz}, when $H(x,\zeta,t)$ and  $E(x,\zeta,t)$ are plotted as a function of $\zeta$ with fixed value of the fraction of momentum carried by the active quark $x$, the peak occurs at same value of $\zeta$. However, the amplitude decreases as the value of $|t|$ increases. Since $\zeta$ represents the longitudinal momentum transfer in the process and we are working in the region $\zeta < x < 1$, therefore, $H(x,\zeta,t)$ increases as the value of the momentum transfer $\zeta$ increases and after reaching a maxima it starts decreasing. It is observed that even if we increase $|t|$, the peak still occurs at same value of $\zeta$. This is due to the fact that the actual momentum during the process is carried by the quark and since it is fixed in this case, therefore any increase in $\zeta$ simply implies the increase in longitudinal momentum fraction. This does not affect the position of the peak which is observed at the same value of $\zeta$ even for different values of $t$.

\section{GPDs in position space}

\begin{figure}
\minipage{0.42\textwidth}
    \includegraphics[width=5.8cm,angle=270]{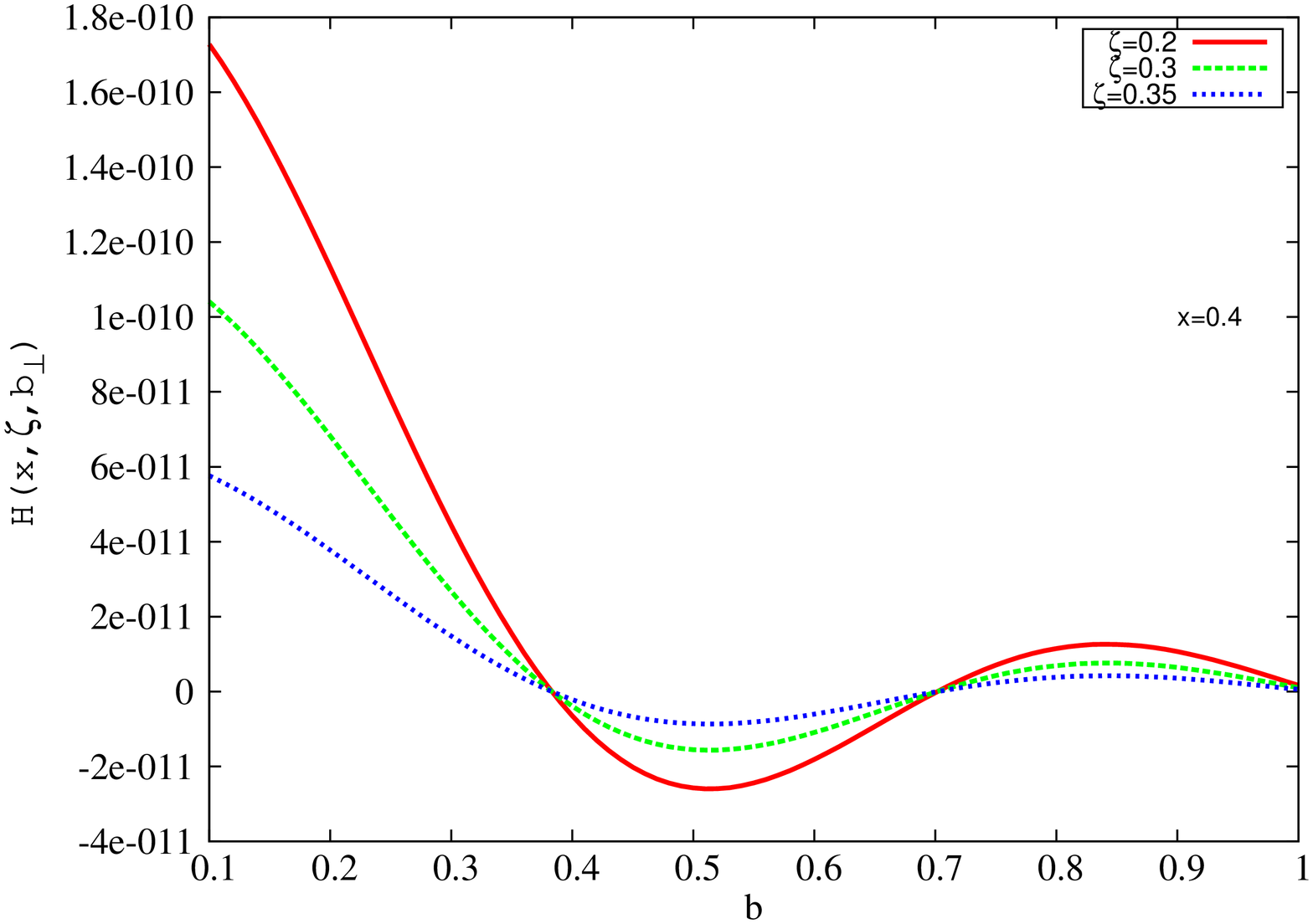}
  \endminipage\hfill
  \minipage{0.42\textwidth}
  \includegraphics[width=5.8cm,angle=270]{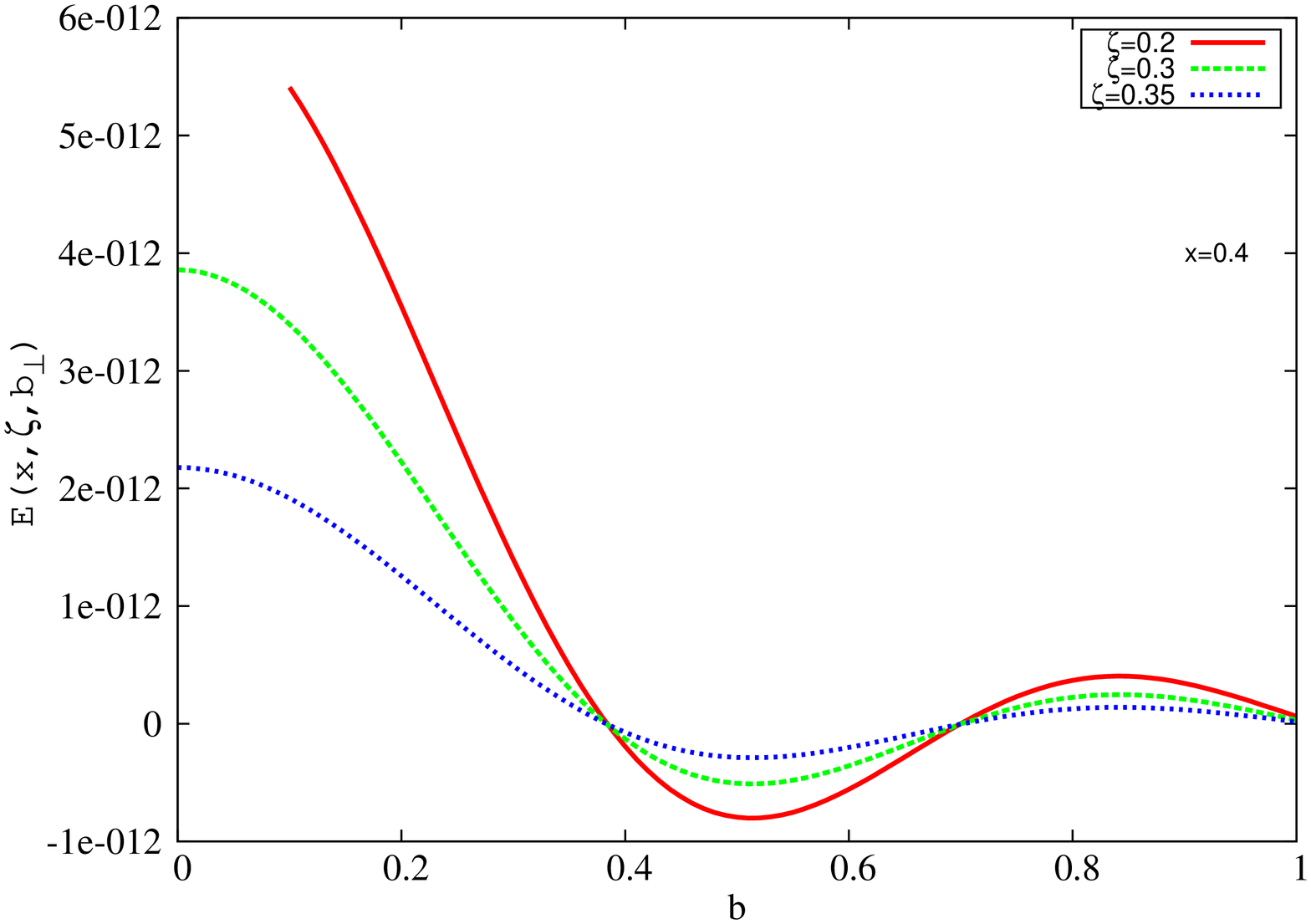}
\endminipage\hfill
\caption{Fourier spectrum of $\mathcal{H}(x,\zeta,b_\perp)$ vs $|b_\perp|$ for fixed value of $x$ and different values of $\zeta$ and Fourier spectrum of $\mathcal{E}(x,\zeta,b_\perp)$ vs $|b_\perp|$ for fixed value of $x$ and different values of $\zeta$.}
\label{newimp}
\end{figure}

\begin{figure}
\minipage{0.42\textwidth}
    \includegraphics[width=5.8cm,angle=270]{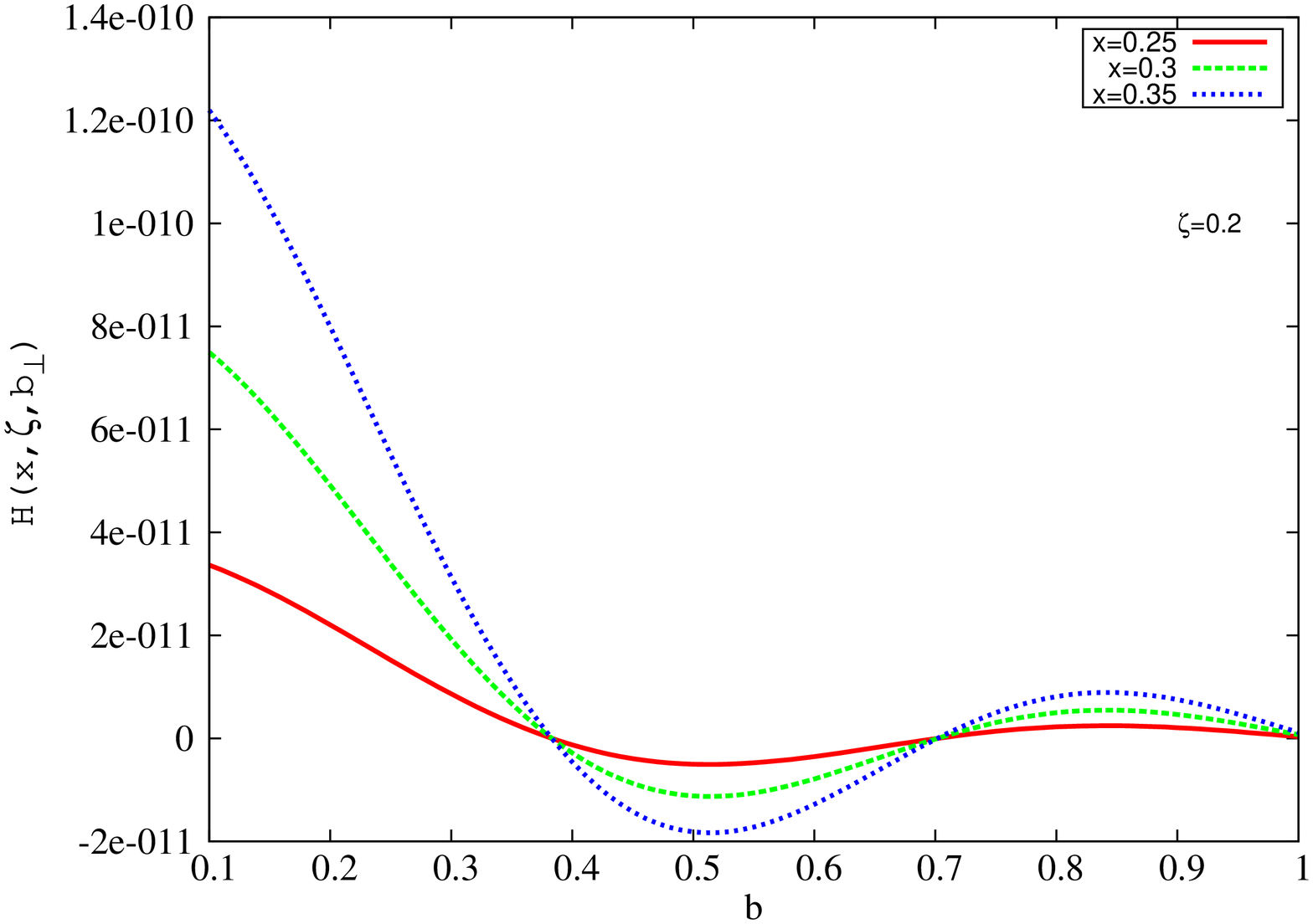}
  \endminipage\hfill
  \minipage{0.42\textwidth}
  \includegraphics[width=5.8cm,angle=270]{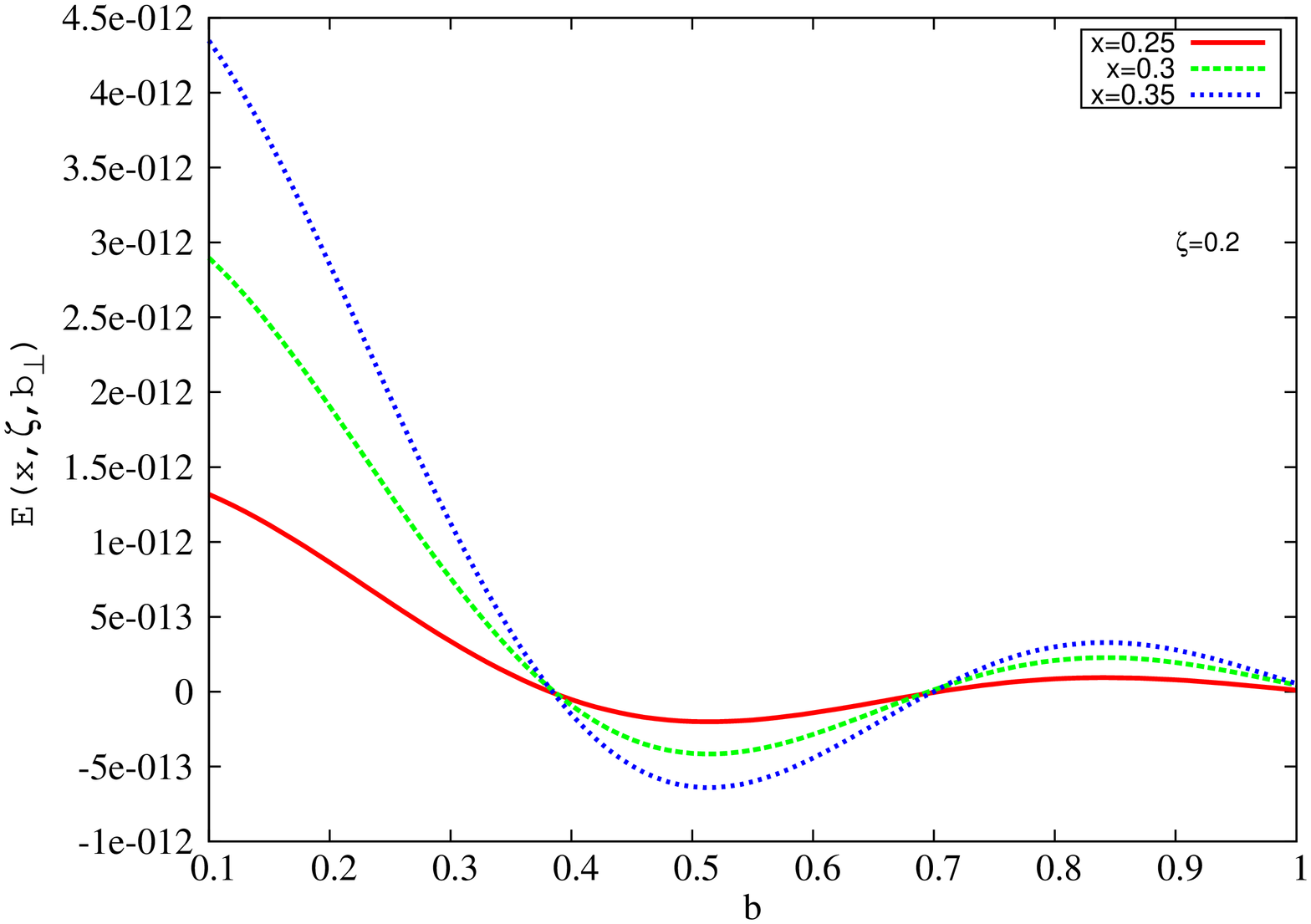}
\endminipage\hfill
\caption{Fourier spectrum of $\mathcal{H}(x,\zeta,b_\perp)$ vs $|b_\perp|$ for fixed value of $\zeta$ and different values of $x$ and Fourier spectrum of  $\mathcal{E}(x,\zeta,b_\perp)$ vs $|b_\perp|$ for fixed value of $\zeta$ and different values of $x$.}
\label{newimpz}
\end{figure}

The FT with respect to the transverse momentum transfer $\Delta_\perp$ gives the GPDs in transverse impact parameter space. We introduce $b_{\perp}$ conjugate to $\Delta_{\perp}$ giving
\bea
{\mathcal H}(x,\zeta,b_{\perp}) &=& \frac{1}{(2\pi)^2} \int d^2 \Delta_\perp  e^{-i \Delta_\perp \cdot b_\perp} H(x,\zeta,t) \,, \nonumber\\
&=&\frac{1}{2\pi} \int \Delta \ d\Delta \ J_0(\Delta b) H(x,\zeta,t)\,,
\eea
\bea
{\mathcal E}(x,\zeta,b_{\perp}) &=& \frac{1}{(2\pi)^2} \int d^2 \Delta_\perp  e^{-i \Delta_\perp \cdot b_\perp} E(x,\zeta,t) \,, \nonumber\\
&=&\frac{1}{2\pi} \int \Delta \ d\Delta \ J_0(\Delta b) E(x,\zeta,t)\,,
\eea
where $\Delta$=$|\Delta_\perp|$. Here $b$=$|b_\perp|$ is the impact parameter measuring the transverse distance between the struck parton and the center of momentum of the hadron and is defined as $\sum x_i b_i = 0$ (the sum being over the number of partons). Since we are working in the limit of non-zero skewness, the transverse location of the proton before and after scattering is different and this transverse shift depends upon $\zeta$ and $b_\perp$. Even when integrated over $x$ the information does not wash away.  In the DGLAP region ($\zeta < x < 1$), the parameter $b=|b_\perp|$ gives the location of the quark where it is pulled out and put back to the nucleon whereas in ERBL region it describes the location of quark-antiquark pair inside the nucleon.
In Fig. \ref{newimp}, we  plot $\mathcal {H}(x,\zeta,b_{\perp})$ and $\mathcal {E}(x,\zeta,b_{\perp})$ as a function of $|b_\perp|$. We have fixed the value of $x$ and have taken different values of $\zeta$. We know that $\zeta$ can never be zero in the experiments because their is always a finite momentum transfer in the longitudinal direction. Therefore, probability interpretation is not possible.
It is clear from the plots that the peak of GPDs in transverse impact parameter space decreases as value of $\zeta$ increases.
For the sake of completeness, in Fig. \ref{newimpz},  we have plotted $\mathcal {H}(x,\zeta,b_{\perp})$ and $\mathcal {E}(x,\zeta,b_{\perp})$ as a function of $|b_\perp|$. In this case however, we have fixed the value of $\zeta$ and have taken different values of $x$.
In contrast to Fig. \ref{newimp}, in Fig. \ref{newimpz}, the peak of GPDs in transverse impact parameter space increases $x$ increases. The smearing in the $b_\perp$ space is due to the multiparticle correlation. As we increase the value of $\zeta$ for a fixed value of $x$, smearing in the $b_\perp$ space increases implying that the momentum transfer in the longitudinal direction increases.

Since we are effectively looking at the initial and final waves of the scattered proton where the final state proton wave function gets modified to the incident proton wave function because of the momentum transferred to the quark in Compton scattering. The change in quark momentum along the longitudinal direction can be Fourier transformed to a shift in the light front position of the struck quark. Similar to the transverse momentum transfer, following Ref. \cite{brodsky3,brodsky4}, we introduce a longitudinal boost invariant impact parameter $\sigma$, which is conjugate to longitudinal momentum transfer $\zeta$. One can thus simulate a change in the quark's longitudinal light front co-ordinate by an amount $\sigma=\frac{1}{2} b^- P^+$, designated as the scaling parameter. This is analogous to diffractive scattering of a wave in optics where $\sigma$ plays the role of the physical size of the scattering center in a one-dimensional system. We are using $t$ to set the change in the transverse momentum of the quark in the scattering the scale used is dynamical in nature.
The GPDs in longitudinal position space are now expressed as
\bea
\mathcal {H}(x,\sigma,t)=\frac{1}{2\pi} \int_{0}^{\zeta_{max}} d\zeta \ e^{i  \frac{1}{2}  P^+  \zeta  b^-} H(x,\zeta,t)\,,
\eea
\bea
\mathcal {E}(x,\sigma,t)=\frac{1}{2\pi} \int_{0}^{\zeta_{max}} d\zeta \ e^{i  \frac{1}{2}  P^+  \zeta  b^-} E(x,\zeta,t)\,.
\eea
\begin{figure}
\begin{center}
\minipage{0.42\textwidth}
 \includegraphics[width=5.8cm,angle=270]{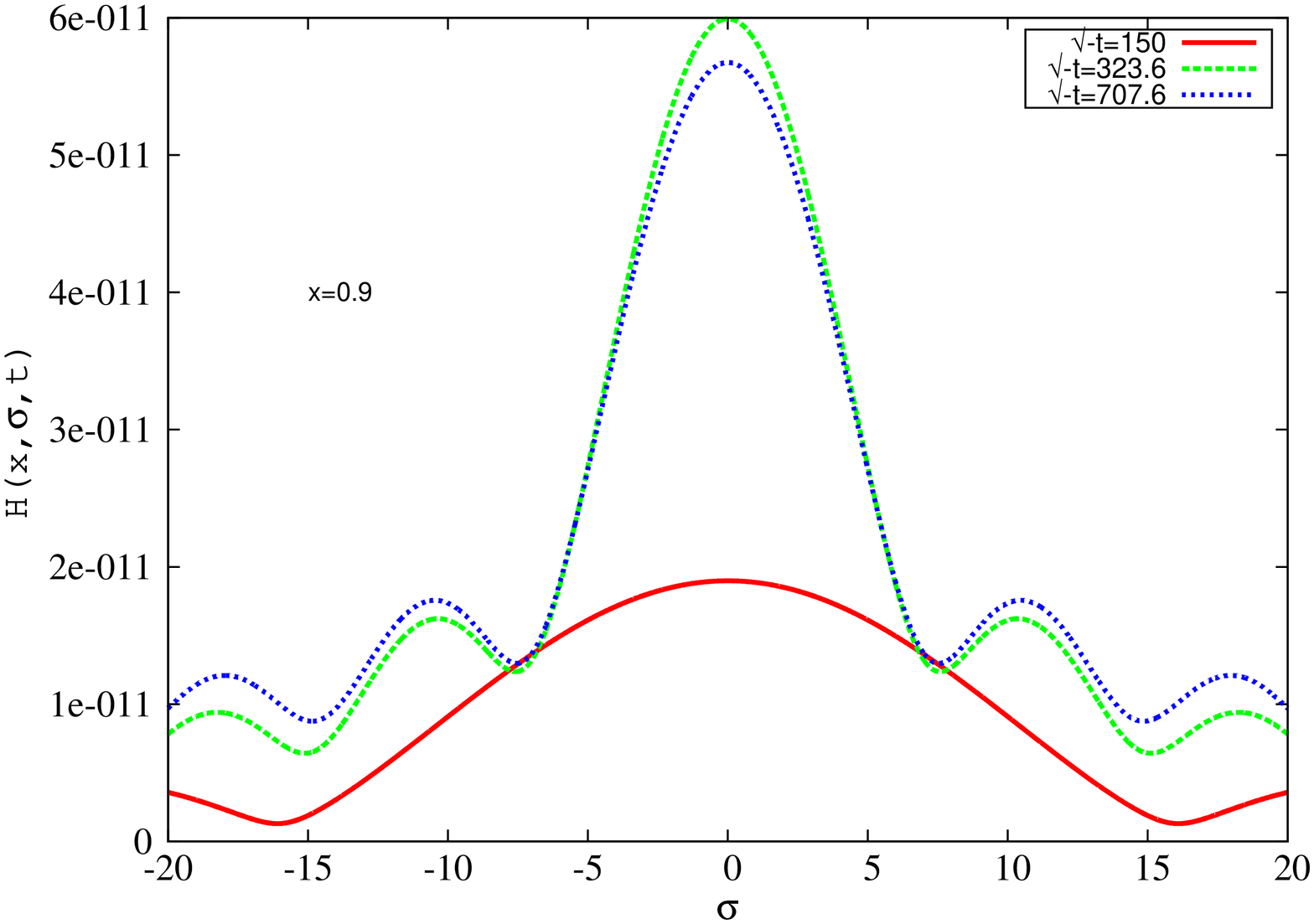}
  \endminipage\hfill
  \minipage{0.42\textwidth}
  \includegraphics[width=5.8cm,angle=270]{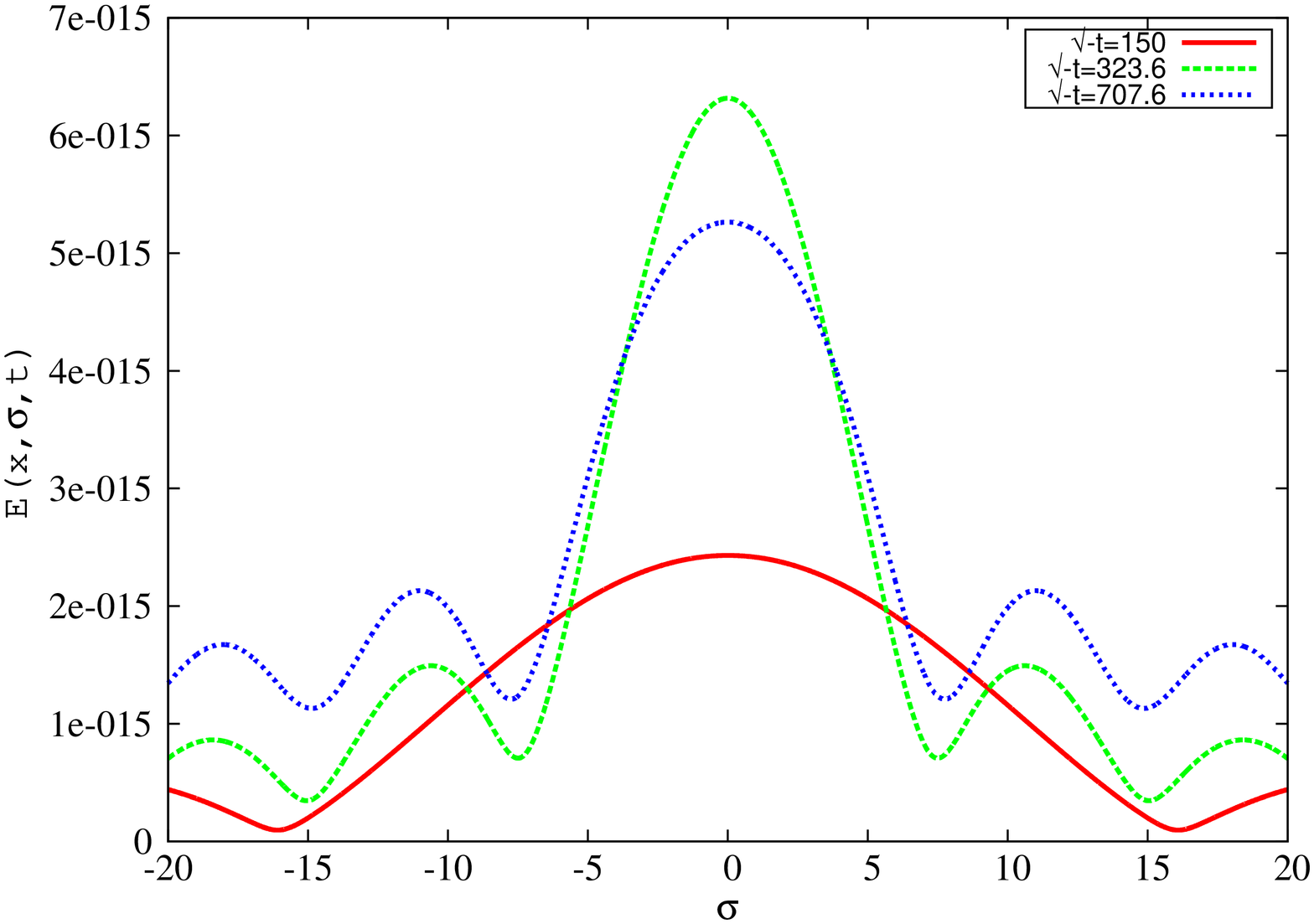}
    \endminipage\hfill
    \caption{Fourier spectrum of $\mathcal{H}(x,\sigma,t)$ vs $\sigma$ for
fixed value $x$ with different values of $-t$ and Fourier spectrum of
$\mathcal{E}(x,\sigma,t)$ vs $\sigma$ for  fixed value $x$ with different values of $-t$. Parameter $t$ is in $MeV^2$} \label{newf}
\end{center}
\end{figure}
In the region $\zeta<x<1$, the upper limit of integration is given by $\zeta_{max}$
\be
\zeta_{max}=\frac{-t}{2 M^2}(\sqrt{1+\frac{4 M^2}{(-t)}}-1)\,,
\ee
whereas in the region $0<x<\zeta$, the upper limit is given by the value of $x$.

In Fig. \ref{newf}, we plot the Fourier pattern for $\mathcal {H}(x,\sigma,t)$ and $\mathcal {E}(x,\sigma,t)$ as a function of $\sigma$ in
the longitudinal position space. Diffraction pattern is obtained for both $\mathcal {H}(x,\sigma,t)$ and $\mathcal {E}(x,\sigma,t)$ where
$\zeta_{max}$ plays the role of the slit width. Since the position of the minima (measured from the center of the diffraction pattern) are inversely proportional to the slit width, the minima move away from the center as the slit width ($\zeta_{max}$) decreases. In both the plots the primary maxima is followed by secondary maxima and as the value of $|t|$ increases, the minima moves towards smaller value of $\sigma$. Therefore, the longitudinal size of parton distribution becomes longer and the shape of the conjugate light cone momentum distribution becomes narrower with increasing $|t|$. It can also be added that the position of minima is independent of helicity and it appears at same points in both cases.
 This is a characteristic of diffraction pattern obtained in single slit experiment.

\section{Conclusions}
In the present work we have studied the GPDs in transverse and
longitudinal position space.  We have used a two body model  which
represents a composite system consisting of a fermion and vector boson
with arbitrary masses. The calculations have been done with non-zero
$\zeta$ in the DGLAP region $x>\zeta$. We have considered the
$n\rightarrow n$ parton number conserving overlap. Differentiating the
denominator of the light front wave function w.r.t $M^2$ improves the
convergence near the end points of $x$ and an improved behaviour is observed.

The behaviour of the helicity non-flip GPD $H(x,\zeta,t)$ and helicity flip GPD $E(x,\zeta,t)$ as a function of $x$ with fixed value of $\zeta$ increases with $x$, reaches maximum and then decreases. The active quark carries all the momentum in the limit $x \rightarrow 1$ and the contribution from other partons is expected to be zero. The active quark is more likely to have a large momentum fraction as the value of $|t|$ increases. We are considering here only the leading Fock space component of the target hadron state. The higher Fock space components are more likely to contribute in the small $x$ region.
On the other hand, when $H(x,\zeta,t)$ and  $E(x,\zeta,t)$ are plotted as a function of $\zeta$ with fixed value of the fraction of momentum carried by the active quark $x$, the peak occurs at same value of $\zeta$. This is due to the fact that the actual momentum during the process is carried by the quark and since it is fixed in this case, therefore any increase in $\zeta$ simply implies the increase in longitudinal momentum fraction. This does not affect the position of the peak which is observed at the same value of $\zeta$ even for different values of $t$.

The impact parameter space representations are obtained by taking the Fourier transform of the GPDs with respect to transverse momentum transfer and it probes the parton distributions in impact parameter space. If $|b_\perp|$ is the impact parameter measuring the transverse distance between the struck parton and the center of momentum of the hadron, the transverse location of the proton before and after scattering is different and this transverse shift depends upon $\zeta$ and $b_\perp$. The smearing in the $b_\perp$ space occurs due to the multiparticle correlation and increases with the increase of $\zeta$ implying that the momentum transfer in the longitudinal direction increases.

Further, if we introduce a longitudinal boost invariant impact parameter $\sigma$
conjugate to longitudinal momentum transfer $\zeta$, after Fourier
transform, we get the GPDs in longitudinal position space. Both $H$
and $E$ show the diffraction pattern analogous to diffractive
scattering of a wave in optics where the distribution is in $\sigma$
space. Similar pattern has been observed in some other models also but the general features of this pattern are independent of specific models and depends mainly on the finiteness of $\zeta$ integration as well as the dependence of GPDs on $x$, $\zeta$ and $t$. However, in order to get the full Lorentz invariant picture in longitudinal space, one has to study the GPDs in $x<\zeta$ region as well.

\section{Acknowledgement}
Authors would like to thank S.J. Brodsky and O.V. Teryeav for helpful discussions. HD would like to thank Department of Science and Technology (Ref No. SB/S2/HEP-004/2013), Government of India for financial support. NK would like to thank BLTP, JINR (Russia) for financial support.

\end{document}